\documentstyle[12pt]{article}
\title{\bf Dark energy from conformal symmetry breaking}
\author{
  F. Darabi\thanks{e-mail: f-darabi@azaruniv.edu}
\\
{\small Department of Physics, Shahid Madani University, Tabriz,
53714-161,  Iran.}}
\begin{document}
\maketitle
\begin{abstract}
The breakdown of conformal symmetry in a conformally invariant scalar-tensor
gravitational model is revisited in the cosmological context. Although the old scenario of conformal symmetry breaking in cosmology containing scalar field has already been used in many earlier works, it seems that no special attention has been paid for the investigation on the possible connection between the breakdown of conformal symmetry and the existence of dark energy. In this paper, it is shown that the old scenario of conformal symmetry breaking in cosmology, if properly interpreted, not only has a potential ability to describe the origin of dark energy as a symmetry breaking effect, but also may resolve the coincidence problem. 
\end{abstract}
\vspace{2cm}
\section{Introduction}

Recent cosmological observations obtained by SNe Ia {\cite{c1}}, WMAP {\cite{c2}}, SDSS {\cite{c3}} and X-ray {\cite{c4}} indicate that our universe is experiencing
an accelerated expansion. These observations almost confirm that our
universe is spatially flat, and consists of about $70 \%$ dark
energy (DE) with negative pressure, $30\%$ dust matter (cold dark
matter plus baryons), and negligible radiation. To explain why the cosmic acceleration happens, many theories and models have been proposed. One of
the simplest candidate for the dark energy is a tiny positive cosmological constant $\Lambda$. However, this scenario suffers from cosmological constant and the coincidence problems \cite{4}.
An alternative proposal to explain the dark energy is the dynamical dark energy scenario. The effective dynamical nature of dark energy can originate from various fields, such as canonical scalar field (quintessence)
\cite{quint}, phantom field \cite{phant}, or the combination
of quintessence and phantom in a unified model named quintom
\cite{quintom}. More recently, another theory has been constructed in
the light of the holographic principle of quantum gravity theory which may simultaneously provide a solution to the coincidence problem
\cite{holoprin}.

On the other hand, extended theories of gravity  \cite{mauro, odintsov, farasot, libroSaFe,defelice} try to address the problem of acceleration of the universe. The idea to extend Einstein's theory of gravitation is economic in compared with several attempts which try to solve the above problem by adding new and often unjustified matter ingredients giving rise to clustered structures (dark matter) or to accelerated dynamics (dark energy), thanks to exotic equations of state. The general paradigm of extended theories of gravity consists in adding, into the effective action, physically motivated higher order curvature invariants and non-minimally coupled scalar fields \cite{odi,farh}.
In particular, relaxing the hypothesis that gravitational Lagrangian has to be a linear function of the Ricci scalar $R$, one can take into account, as a minimal extension, an effective action where the gravitational Lagrangian is a generic $f(R)$ function. One can show that this action dynamically corresponds to an action of non-minimally coupled gravity with a new scalar field of
Brans-Dike type having no kinetic term $\omega=0$ known as O'Hanlon action in metric formalism \cite{farasot, Sotiriou}.

It is well known that, conformal symmetry is playing a particularly important role in the investigation of gravitational models including scalar fields. Moreover, the idea of conformal symmetry breaking in cosmology for scalar fields was used many times, see for instance \cite{Grib} about particle creation due to violation of conformal symmetry. In the presence of dimensional parameters, the conformal symmetry can be established for a large class of theories provided the dimensional parameters are conformally transformed according to their dimensions \cite{Bekenstein}. One general feature of conformally invariant theories is the presence of varying dimensional coupling constants. Therefore, the introduction of a constant dimensional parameter into
a conformally invariant theory breaks the conformal symmetry in the sense that a preferred conformal frame, in which the dimensional parameter has
the assumed constant configuration, is singled out. In a conformally invariant gravitational model one usually considers the symmetry breaking 
as a cosmological effect. This would mean that conformal symmetry may be
broken down by defining a preferred conformal frame in terms of the large scale properties of cosmic matter distributed in a finite universe.
In this way, the breakdown of conformal symmetry becomes a framework in which one can look for the origin of the gravitational coupling of matter and a
cosmological constant in large cosmological scales. This goal has been first
achieved by Deser \cite{Deser} and the idea behind it was followed by many people. The purpose of the present paper is to show that the old idea of
cosmological breakdown of conformal invariance in a conformally invariant gravitational model has new features which, if properly interpreted, may be used to explain the accelerating behavior of the universe in matter dominant era and provide a solution to the coincidence problem.

\section{Conformal symmetry and its breakdown in gravitational systems}

This section is devoted to a brief review of the work done by Deser in \cite{Deser}. Consider
the action 
\begin{equation}
S[\phi]=\frac{1}{2} \int \!d^4 x \sqrt{-g} (g^{\mu \nu} \partial_{\mu} \phi
\partial_{\nu} \phi +\frac{1}{6} R \phi^2), 
\label{1}
\end{equation}
which describes a gravitational system, with scalar curvature $R$, consisting of a non-minimally coupled real scalar field $\phi$. Variations with respect
to $\phi$ and $g_{\mu \nu}$ lead to the following equations
\begin{equation}
(\Box -\frac{1}{6} R)\phi=0,
\label{2}
\end{equation}
\begin{equation}
G_{\mu \nu}=6\phi^{-2} \tau_{\mu \nu}(\phi),
\label{3}
\end{equation}
where $G_{\mu \nu}=R_{\mu \nu}-\frac{1}{2}g_{\mu \nu} R$ is the Einstein tensor
and
\begin{equation}
\tau_{\mu \nu}(\phi)= - [\nabla_\mu \phi \nabla_\nu \phi - \frac{1}{2}g_{\mu \nu}
\nabla_\alpha \phi \nabla^\alpha \phi] -
\frac{1}{6}(g_{\mu \nu}\Box -\nabla_\mu \nabla_\nu)\phi^2, 
\end{equation}
with $\nabla_\mu$ denoting the covariant derivative. Taking the trace
of (\ref{3}) gives
\begin{equation}
\phi(\Box -\frac{1}{6} R)\phi=0, 
\label{5}
\end{equation}
which is consistent with equation (\ref{2}). This is a direct consequence of the conformal symmetry of action (\ref{1}) under the conformal transformations
\begin{equation}
\phi \rightarrow \bar{\phi}=\Omega^{-1}(x) \phi ,\:\:\:\:\:\:\:\:\:\:
g_{\mu \nu}\rightarrow \bar{g}_{\mu \nu}=\Omega^2 (x) g_{\mu \nu}, 
\label{6}
\end{equation}
where the conformal factor $\Omega(x)$ is an arbitrary, smooth
function of space-time. Adding a matter source $S_{m}$ to (\ref{1}) gives a generic standard model action as
\begin{equation}
S = S[\phi] + S_{m}, 
\label{7}
\end{equation}
yields the field equations
\begin{equation}
(\Box -\frac{1}{6} R)\phi=0, 
\label{8}
\end{equation}
\begin{equation}
G_{\mu \nu}=6\phi^{-2}[\tau_{\mu \nu}(\phi)+T_{\mu \nu}], 
\label{9}
\end{equation}
where $T_{\mu \nu}$ is the matter energy-momentum tensor which is independent of $\phi$. Then, the following algebraic requirement
\begin{equation}
T_\mu ^{\mu}=0, 
\label{10}
\end{equation}
emerges by comparing the trace of (\ref{9}) with (\ref{8}). This implies that only traceless matter can couple consistently to such conformal invariant
gravitational models. 

The conformal symmetry is broken down by adding a dimensional mass 
term $\frac{1}{2}\int\!d^4 x \sqrt{-g} \mu^2 \phi^2$, with 
$\mu$ being a constant mass parameter, to the action
(\ref{7}). In fact, a conformal transformation requires all dimensional parameters to be transformed according to their dimensions so that $\mu$ should obey the transformation rule $\mu \rightarrow \Omega^{-1}(x) \mu$.
Conformal invariance can be broken down when a particular conformal frame is chosen in which the dimensional parameter (like $\mu$) takes on a constant configuration. The choice of such a specific conformal frame is usually suggested
by the physical considerations one wishes to investigate in the problem at hand. Considering the presence of the new mass term in the action leads to
\begin{equation}
\mu^2 \phi^2=T_\mu ^{\mu}.
\label{12}
\end{equation}
Consequently, the field equations become
\begin{equation}
(\Box -\frac{1}{6} R-\mu^2)\phi=0, 
\label{13}
\end{equation}
\begin{equation}
G_{\mu \nu}-3\mu^2 g_{\mu \nu}=6\phi^{-2} [\tau_{\mu \nu}(\phi)+T_{\mu \nu}].
\label{14}
\end{equation}
Here, we intend to determine a conformal frame using the large scale properties of the observed universe. This would mean that 
one may take $\mu^{-1}$ as the length scale characterizing the 
typical size of the universe $a_0$ and $T_\mu^\mu$ 
as the average density of the large scale distribution
of matter $ \sim M a_0^{-3}$, $M$ being the observed mass of the universe.
This leads, as a consequence of (\ref{12}) to the estimation of the 
constant background value of $\phi$ as follows
\begin{equation}
{\phi}^{-2} \sim a_0^{-2}(M/a_0^3)^{-1} \sim a_0/M \sim G,
\label{15}
\end{equation}
where the well-known empirical cosmological relation $GM/a_0 \sim 1$
is used. Inserting this background value of $\phi$ into the field
equations (\ref{13}), (\ref{14}) leads to the following set of Einstein equations
\begin{equation}
G_{\mu \nu}-3\mu^2 g_{\mu \nu}=6{\phi}^{-2}
T_{\mu \nu} = 8\pi G T_{\mu \nu} ,
\label{16}
\end{equation}
with a correct coupling constant $8\pi G$, and a
term $3\mu^2$ which appears as an effective cosmological constant $\Lambda$
of the order of $ a_0^{-2}$. 
Note that no new information is contained in the field equation (\ref{13}) for $\phi$. In conclusion, we have obtained a preferred conformal frame $(\phi,g_{\mu\nu},\mu)$ for the gravitational variables, where $\phi^2\sim G^{-1}$, 
$\mu^2\sim\Lambda$ and $g_{\mu\nu}$ is determined by the
field equations (\ref{16}). This preferred confromal frame has the remarkable
property that shows a correct coupling of the cosmic matter to gravity, so we shall call it the cosmological frame.

\section{Dynamics of the universe before and after symmetry breaking}

\subsection{Radiation dominance}

Let us assume the action of our cosmological model in the radiation dominant
era as
\begin{equation}
S = S[\phi] + S_{r}, 
\label{7'}
\end{equation}
where $S_{r}$ is the action corresponding to the radiation. We take $g_{\mu \nu}$ as the flat ($k=0$) Robertson-Walker metric
\begin{equation}
ds^2=-dt^2+a^2(t)\left[dr^2+r^2(d\theta^2+\sin^2\theta
d\phi^2 )\right], \label{22}
\end{equation}
and $T^{~r}_{\mu \nu}$ as perfect fluid describing the radiation
\begin{equation}
{T}^{~r}_{\mu \nu}=(\rho_{r}+p_{r}){u}_{\mu} {u}_{\nu}+p_{r}{g}_{\mu \nu}, \label{23}
\end{equation}
where $a$ is the scale factor, $\rho_{r}$ is the density and $p_{r}$
is the pressure of the radiation. It is easy to show that the dominant energy-momentum tensor in this era is traceless, $T_\mu ^{\mu~r}=0$. Therefore,
according to the arguments in the previous section, the conformal symmetry holds dominantly in this stage of universe evolution. 
The field equations (\ref{8}), (\ref{9}) are now explaining a cosmology of scalar-tensor type. Substituting $g_{\mu \nu}$ and $T_{\mu \nu}^{~r}$ into these equations
yields
\begin{equation}
\frac{\dot{a}^2}{a^2}+\frac{\dot{\phi}^2}{\phi^2}+2\frac{\dot{a}}{a}\frac{\dot{\phi}}{\phi}=2\frac{\rho_{r}}{\phi^2},\label{24'}
\end{equation}
\begin{equation}
2\frac{\ddot{a}}{a}+\frac{\dot{a}^2}{a^2}=-6\frac{p_{r}}{\phi^2},\label{25'}
\end{equation}
\begin{equation}\label{26}
\frac{\ddot{\phi}}{\phi}+\frac{\ddot{a}}{a}+3\frac{\dot{a}}{a}\frac{\dot{\phi}}{\phi}
+\frac{\dot{a}^2}{a^2}=0,
\end{equation}
where $\dot{a}$ means time derivative with respect to $t$.
By combining Eqs.(\ref{24'}) and (\ref{25'}) we obtain the acceleration
equation
\begin{equation}
\frac{\ddot{a}}{a}=\frac{\dot{\phi}^2}{2{\phi}^2}+\frac{\dot{a}}{a}\frac{\dot{\phi}}{\phi}
-\frac{1}{{\phi}^2}(\rho_{r}+3p_{r}),\label{26'}
\end{equation}
and the conservation equation
\begin{equation}
\dot{\rho}_r=\frac{\dot{a}}{a}\left[\frac{1}{2}\frac{\dot{\phi}^2}{{\phi}^2}+2\frac{\dot{a}}{a}\frac{\dot{\phi}}{\phi}
-3\frac{(\rho_{r}+p_{r})}{{\phi}^2}+\frac{\ddot{\phi}}{\phi}\right]{\phi}^2. \label{27}
\end{equation}
If we put the power law behaviors $\rho_{r}=A a^{\alpha}$, $\phi=B a^{\beta}$
and $a=Ct^{\gamma}$ together with the equation of state $p_{r}=\frac{1}{3} \rho_{r}$ (for radiation dominant era) into Eq.(\ref{27}) we obtain the following equations
\begin{equation}
\gamma=\frac{2}{2\beta-\alpha},\label{28}
\end{equation}
\begin{equation}
\alpha A-2\beta B^2\frac{C^{2\beta-\alpha}}{2\beta-\alpha}\left[\frac{2}{2\beta-\alpha}\left(\frac{3\beta}{2}+2\right)-1\right]=\frac{6A}{B^2},\label{29}
\end{equation}
\begin{equation}
2\frac{C^{2\beta-\alpha}}{2\beta-\alpha}\left[\frac{2}{2\beta-\alpha}-1-\frac{2\beta}{2\beta-\alpha}\left(\frac{\beta}{2}+1\right)-1\right]=\frac{2}{B^2}.\label{30}
\end{equation}
It is easy to see that the universe may experience decelerating phase in
radiation dominant era provided that 
\begin{equation}
\beta>1+\frac{\alpha}{2}.\label{31}
\end{equation}
Therefore, for some given suitable set of initial conditions $\{A,B,C\}$, one can obtain, through Eqs.(\ref{28})-(\ref{30}), the values of $\alpha$, $\beta$, and $\gamma$ for which the above inequality is satisfied and we have a decelerating behavior for the radiation dominant era in the context of present scalar-tensor cosmology \footnote{Note that, one may reconstruct the radiation dominant era in the standard cosmology (broken phase of conformal symmetry with a constant scalar field $\phi$) by the following choices
\begin{equation}
\alpha=-4,~~~~~~~\beta=0, ~~~~~~~\gamma=\frac{1}{2}.
\end{equation}
}. 

\subsection{Matter dominance}

The decelerating behavior of the universe continues until the universe experience a phase transition from radiation dominant to matter dominant eras. By this,
we merely mean a phase transition in the cosmological context where a dominant
source of radiation with nonzero pressure have been changed into a source of visible matter with vanishing pressure.

When this transition occurs the action (\ref{7'}) is changed, in the first place, as
\begin{equation}
S = S[\phi] + S_{m}, 
\label{7''}
\end{equation}
where $S_{r}$ has been replaced by $S_{m}$ as the dominant contribution. The energy-momentum tensor associated with $S_{m}$ becomes 
\begin{equation}
{T}^{~m}_{\mu \nu}=(\rho_{m}){u}_{\mu} {u}_{\nu}, \label{23'}
\end{equation}
with a non-vanishing trace 
\begin{equation}\label{23''}
T_\mu ^{\mu~m}=\rho_{m}.
\end{equation}
Then, according to the discussion in section 2, the field equations (\ref{8}), (\ref{9}) are not consistent unless a mass term is added to the action (\ref{7''}) as
\begin{equation}
S = S[\phi] + \frac{1}{2}\int\!d^4 x \sqrt{-g} \mu^2 \phi^2+S_{m}, 
\label{7'''}
\end{equation}
which clearly breaks down the conformal symmetry. Note that, the mass term in Deser's theory is included by hand with no physical justification,
or interpretation. Whereas, in the present cosmological framework, the appearance
of mass term in the action is naturally interpreted due to the cosmological dominance of matter over radiation in the matter dominant era. Such a mass
term is not required in radiation dominant era because the energy-momentum
tensor is traceless. Hence, conformal symmetry is not broken in the radiation dominant era and so the dark energy and the resultant acceleration of the universe are not appeared. 

According to this scenario, if we naturally accept that the cosmological conformal symmetry is broken down during the transition from radiation dominance to matter dominance, then an important consequence of this symmetry breaking is the appearance of gravitational coupling $\phi^2\sim G^{-1}$, and a positive cosmological constant $\mu^2\sim\Lambda$ in the Einstein equation which describes
the evolution of the universe in matter dominant era as 
\begin{equation}
G_{\mu \nu}= 8\pi G T_{\mu \nu}+ \Lambda g_{\mu \nu}.
\label{16'}
\end{equation}
Therefore, the induced positive cosmological constant $\Lambda$ resulting
from conformal symmetry breaking in the matter dominant era becomes a potential candidate for dark energy which is responsible for the acceleration of the universe.

\section{Coincidence problem}

It is known that the matter density of the universe scales with the expansion of the universe as $\rho_m \sim {1}/{a^3}, $ and the vacuum energy density $\rho_V$ which comes from quantum field vacuum fluctuations is almost constant. If we ignore the cosmological constant problem, it seems there is only one epoch in the history of the universe for which it happens that $\rho_V \sim \rho_m$. It is so difficult to understand why we happen to exist in this special epoch. In other words: How finely-tuned is it that we live in the special era where the vacuum and matter densities are comparable? This is known as {\it Coincidence problem} \cite{Carr}.
In this section, we aim to show that the present scenario is also capable
of solving the coincidence problem. We know the energy density of the vacuum is given by
\begin{equation}
\rho_V=\frac{\Lambda}{8 \pi G},\label{33}
\end{equation}
and according to the model of Deser we have 
\begin{equation}
\rho_V\sim \frac{\mu^2}{G}.\label{34}
\end{equation}
On the other hand, using $\phi^2\sim G^{-1}$, (\ref{12}) and (\ref{23''}) we obtain
\begin{equation}\label{35}
\mu^2 \sim G\rho_m.
\end{equation}
Combining (\ref{34}) and (\ref{35}) leads to the desired result
\begin{equation}\label{36}
\rho_V \sim \rho_m.
\end{equation}
This shows that the coincidence problem, similar to the dark energy problem, is subject to the cosmological matter dominance and is a direct consequence of the conformal symmetry breaking in the matter dominant era.

\section{Discussion and Conclusions}

The concept of symmetry in physics is of fundamental importance. However,
physics at the fundamental level much benefits of the symmetry breaking effect.
The most important example is the spontaneous symmetry breaking which results
in the emergence of fundamental interactions. In this paper, motivated by this general feature of nature, we have shown that the model of Deser for the breakdown of conformal symmetry in a gravitational system may be used to explain the emergence of dark energy and acceleration of the universe
in the matter dominant era, as well as the coincidence problem. This is done by assuming a scalar-tensor cosmology before symmetry breaking and the standard cosmology after symmetry breaking.

In the scalar-tensor sector with radiation dominance, we have a conformally invariant gravitational system coupled to a massless scaler field 
$$
S = S[\phi] +\underbrace{\frac{1}{2}\int\!d^4 x \sqrt{-g} \mu^2 \phi^2}_{vanishing}+S_{r},
$$
where the traceless condition $T_\mu ^{\mu}=0$ implies $\mu^2=0$, and $\phi$ plays the role of dynamical gravitational coupling. The field equations show
the possibility of having a decelerating universe. 

On the other hand, in the matter dominance, we have $T_\mu ^{\mu}\neq0$ which leads to the breakdown
of cosmological conformal symmetry by inducing a non-vanishing mass term $$
S = S[\phi] + \frac{1}{2}\int\!d^4 x \sqrt{-g} \mu^2 \phi^2+S_{m}.
$$
Then, a constant gravitational coupling as well as a positive cosmological constant , playing the role of dark energy, are emerged through a constant background value of $\phi$, and the standard field equations describe an accelerating universe. 

It is not hard to generalize this scenario to include a conformal invariant
scalar potential and redo this investigation. 
For example, one may consider the action
$$
S[\phi]=\frac{1}{2} \int \!d^4 x \sqrt{-g} (g^{\mu \nu} \partial_{\mu} \phi
\partial_{\nu} \phi +\frac{1}{6} R \phi^2-\lambda \phi^4), 
$$
where $\lambda$ is an small dimensionless self-interacting coupling. The modifications due to this self interacting potential is as follows. In Eq.(\ref{2}) the derivative of this potential, and in Eq.(\ref{3}) (in the definition of $\tau_{\mu \nu}$) the potential itself as $\lambda \phi^4 g_{\mu \nu}$, are appeared as extra terms in the right hand sides, respectively. The modified equations become consistent with each other provided the trace of the latter equation is compared with the former equation. This consistency is a sign of the conformal symmetry. One may write down the field equations
in this conformal invariant phase, where $T_\mu ^{\mu}=0$, and use the power
law solutions for radiation dominant era. It is then possible to arrange
the parameters for occurring the decelerating phase in radiation dominant era, as is done in section 3.1.

The corresponding conformal symmetry is then broken by introducing the mass term $\frac{1}{2}\int\!d^4 x \sqrt{-g} \mu^2 \phi^2$ which, as in Eq.(\ref{12}), ends up with $\mu^2 \phi^2=T_\mu ^{\mu}$. This mass term causes the above
modified conformal invariant equations to take mass terms including $\mu^2$.
These equations, namely Eqs.(\ref{13}), (\ref{14}) with extra terms due to the self interacting potential, are not conformal invariant.
In the special cosmological frame defined by (\ref{15}) we obtain the modified
Einstein equation
$$
G_{\mu \nu}-(3\mu^2+\frac{36 \lambda}{8\pi G}) g_{\mu \nu}= 8\pi G T_{\mu \nu},
$$
where the modification term $({36 \lambda}/{8\pi G}) g_{\mu \nu}$ contributes to the cosmological constant (dark energy), as a result of the self interacting potential. This investigation shows that the addition of conformaly invariant self interacting potential to the original action (\ref{1}), does not change our interpretation of the dark energy as a conformal symmetry breaking effect. The only difference between the action (\ref{1}) and the above modified action is in the value of the dark energy. In the case of action (\ref{1}), the
dark energy density has a value defined by $\mu^2$ and $G$ 
$$\rho_V=\frac{\mu^2}{8\pi G}.$$
However, in the case of above modified action, the value of dark energy density
is shifted by a constant which is determined by $\mu^2$, $G$ and $\lambda$ as
$$\rho_V=\frac{8\pi G \mu^2+\lambda}{(8\pi G)^2}.$$
We may also reconsider the coincidence problem in this case. To this end,
we rewrite the above equation as
$$
\rho_V = \left(\rho_m+\frac{\lambda}{(8 \pi G)^2}\right).
$$
The smallness of second term shows that the dark energy density is again
of the order of matter density, so the coincidence problem is again resolved.

In conclusion, it seems the addition of conformal invariant terms like $\lambda \phi^4$ to the action (\ref{1}) does not alter our interpretation of dark
energy as a conformal symmetry breaking effect. It just changes infinitesimally the value of cosmological constant (dark energy density), and we can still resolve the coincidence problem within the scenario presented in this paper. 
This model, although describes the late time acceleration of the universe as a consequence of conformal symmetry breaking at matter dominant era, however, it is appealing to investigate whether this model can be generalized in its scalar-tensor sector to incorporate an inflationary phase at early universe. For example, one may think about a specific inflaton (scalar) field with a potential respecting the conformal symmetry. Of course, the extreme conditions in inflationary epoch with ultra high energy of Planck order, in principle, requires quantum gravity or at least quantum field theory. It is also known that the conformal symmetry breaking occurs in these theories via corresponding conformal anomaly and anomaly induced dynamics \cite{Anton}. A conformal anomaly is a quantum phenomenon that breaks the conformal symmetry of the classical theory. So, the conformal symmetry may possibly be broken down in inflationary epoch by a nonvanishing constant
mass term due to the conformal anomaly. Therefore, the generalization of present model to incorporate an inflationary phase should be limited to a pure classical study and not the quantum one. Then, the conformal anomalies become hidden and the conformal symmetry, in principle, may exist in the inflationary epoch as well as radiation dominant era \footnote{In a throughout study of conformal symmetry breaking at early universe, one may think that even if the conformal symmetry would break down in the high energy stage of inflationary epoch via quantum conformal anomalies, it could be restored in the low energy stage of radiation dominant era by some possible mechanisms to make the energy-momentum tensor be classically traceless.  In this regard,
the radiation dominant era represents the conformal symmetry as it should
be.}.

\section*{Acknowledgment}
The author would like to thank the anonymous referee for the enlightening comments.

\end{document}